\documentclass[twocolumn]{aastex62}
\turnoffeditone
\usepackage[english]{babel}
\usepackage{blindtext}
\usepackage{enumitem}
\usepackage{mathtools}
\usepackage[flushleft]{threeparttable}

\graphicspath{{./}{figures/}}

\received{Nov. 27}
\revised{Feb. 06}
\accepted{Feb. 10}
\submitjournal{ApJ}

\shorttitle{Mass Bias of Shear-selected Clusters}
\shortauthors{Chen et al.}

\begin{document}

\title{Mass Bias of Weak Lensing Shear-selected Galaxy Cluster Samples}

\email{kfchen@asiaa.sinica.edu.tw, b04901029@ntu.edu.tw}

\author[0000-0002-3839-0230]{Kai-Feng Chen}
\affiliation{Department of Physics, National Taiwan University}
\affiliation{Department of Mathematics, National Taiwan University}
\affiliation{Institute of Astronomy and Astrophysics Academia Sinica, Taipei 10617, Taiwan}

\author[0000-0003-3484-399X]{Masamune Oguri}
\affiliation{Research Center for the Early Universe,
University of Tokyo, Tokyo 113-0033, Japan}
\affiliation{Department of Physics, University of Tokyo, Tokyo 113-0033, Japan}
\affiliation{Kavli Institute for the Physics and Mathematics of the Universe (Kavli IPMU, WPI), University of Tokyo, Chiba 277-8582, Japan}

\author[0000-0001-7146-4687]{Yen-Ting Lin}
\affiliation{Institute of Astronomy and Astrophysics Academia Sinica, Taipei 10617, Taiwan}

\author[0000-0002-1962-904X]{Satoshi Miyazaki}
\affiliation{National Astronomical Observatory of Japan, 2-21-1 Osawa, Mitaka, Tokyo 181-8588, Japan }
\affiliation{Department of Astronomical Science, The Graduate
University for Advanced Studies (Sokendai), 2-21-1, Osawa,
Mitaka, Tokyo 181-8588, Japan}

\begin{abstract}

We estimate the \added{\edit1{Eddington}} bias on weak lensing mass measurements of shear-selected galaxy cluster samples. The mass bias is expected to be significant because constructions of cluster samples from peaks in weak lensing mass maps and measurements of cluster masses from their tangential shear profiles share the same noise. We quantify this mass bias from large sets of mock cluster samples with analytical density profiles and realistic large-scale structure noise from ray-tracing simulations. We find that, even for peaks with signal-to-noise ratio larger than $4.0$ in weak lensing mass maps constructed in a deep survey with a high source galaxy number density of $30~\mathrm{arcmin}^{-2}$, derived weak lensing masses for these shear-selected clusters are still biased high by $\sim55\%$ on average. Such a large bias mainly originates from up-scattered low mass objects, which is an inevitable consequence of selecting clusters with a noisy observable directly linked to the mass measurement. We also investigate the dependence of the mass bias on different physical and observational parameters, finding that the mass bias strongly correlates with cluster redshifts, true halo masses, and selection signal-to-noise thresholds, but having moderate dependence on observed weak lensing masses and survey depths. This bias, albeit considerable, can still be modeled accurately in statistical studies of shear-selected clusters as the intrinsic scatter around the mean bias is found to be reasonable in size. We demonstrate that such a bias can explain the deviation in X-ray properties previously found on a shear-selected cluster sample. Our result will be useful for turning large samples of shear-selected clusters available in future surveys into potential probes of cosmology and cluster astrophysics.

\end{abstract}

\section{Introduction} 
\label{sec:intro}
Clusters of galaxies have proven to be a sensitive probe of cosmology \citep[for a review, see][]{Allen2011}. The number counts of these gravitationally collapsed structures are, in particular, sensitive to both the geometry and the structure formation history in our Universe. The availability of clusters across a large redshift range allows us to obtain better constraints on dynamic parameters such as the dark energy equation of state compared to analyses relying on a single snapshot of the cosmic history \citep{Weinberg2013}. Modern cluster cosmology constraints are usually derived from complex likelihood analyses \citep{Vikhlinin2009, Mantz2010, Rozo2010, Mantz2014, PlanckSZ, deHaan2016, Bocquet2019}, whose success relies heavily on one's ability to understand the relations between halo masses and various observables from which cluster samples are selected.

Large cluster samples suitable for constraining cosmology are usually selected through observables that are related to baryonic properties of galaxy clusters. Indeed, due to the extreme environment within these massive systems, the swarming galaxies and the hot intracluster medium (ICM) can be detected across a wide range of wavelengths. Modern cluster samples are usually constructed through observations in one of the following three wavelength regimes: with X-ray signals \citep[e.g.,][]{Edge1990, Gioia1990, Vikhlinin1998, Clerc2014} from the thermal Bremsstrahlung; at millimeter wavelengths \citep[e.g.,][]{ACT, SPT, PlanckSZ} as a result of the thermal Sunyaev-Zel'dovich effect \citep{tSZ}; and in the optical via overdensities of red galaxies \citep[e.g.,][]{Gladders2005, Koester2007, CAMIRA, Rykoff2016} or probing clustering directly with the matched filter technique \citep[e.g.,][]{Wen2009, Milkeraitis2010, Szabo2011, AMICO}. These observables are linked to halo masses through semi-empirical scaling relations. Using these baryonic properties alone to constrain free parameters in the scaling relations together with the cosmological parameters, a framework known as \textit{self calibration} \citep{Majumdar2004}, is usually difficult. State-of-the-art cluster cosmology analyses require additional mass information to perform \textit{mass calibration} on these samples \citep[e.g.,][]{Wu2010, OK2011, Huterer2015}.

Weak gravitational lensing is one of the best means to provide an accurate mass estimate \citep[e.g.,][]{Hoekstra2007, Bardeau2007, Hamana2009, Zhang2010, Okabe2010, Mahdavi2013, Okabe2016, Umetsu2016, Dietrich2019}. Measurements of shape distortions of background galaxies, a quantity often referred to as the shear, can be used to derive nearly unbiased projected cluster masses on average \citep{Clowe2004, Corless2007, Becker2011, Bahe2012}. However, cluster masses derived from weak lensing suffer from various sources of scatter and are often of low signal-to-noise ratio ($S/N$). Thus, a large number of clusters with weak lensing mass measurements is still needed to reduce the uncertainties in the mass--observable relations to provide accurate cosmological constraints.

On the other hand, recent wide and deep optical surveys offer an opportunity to construct cluster samples in an alternative way, that is, through identifying peaks in mass maps reconstructed from weak lensing shear maps \citep[e.g.,][]{Wittman2001, Schirmer2007, Miyazaki2007, Shan2012, Miyazaki2015, Miyazaki2019}. Using these shear-selected samples is advantageous for cosmological studies, as they follow a more direct mass--observable relation and their selection functions can be quantified accurately using analytical calculations as well as ray-tracing simulations \citep{Hamana2012, Miyazaki2019}.

The sample of such shear-selected clusters is also useful for understanding the mass--observable relations better, given that they represent a sample of clusters selected solely through their mass distributions and their selection functions are therefore not directly affected by complex baryon physics. For instance\deleted{,} \citet{Giles2015} studied X-ray properties of shear-selected clusters from \citet{Miyazaki2007} to find that shear-selected clusters appeared to be X-ray underluminous for a given mass by a factor of $\sim 2-3$ compared with X-ray selected clusters. A recent paper by \cite{Miyazaki2019}, who constructed a large homogeneous sample of shear-selected clusters from the Subaru Hyper Suprime-Cam (HSC) survey \citep{HSC}, reached a similar conclusion based on analyses of archival X-ray data as well as comparisons with published X-ray catalogs. However, whether these clusters are really X-ray underlumious or their masses are overestimated need to be examined more carefully first. We need to carefully quantify the selection effects of shear-selected clusters and study their impact on the X-ray luminosity--mass relation.

This work serves as a preparatory step for cluster studies utilizing shear-selected clusters by investigating their mass--observable relation under a fixed cosmology. Roughly speaking, we explore the distribution $P(\gamma | M, z)$ where $\gamma$ denotes the observable, namely the weak lensing shear profile, and $M, z$ represent the underlying halo mass and redshift. This distribution can be decomposed into $P(\gamma | M, z) = \int P(\gamma | M_{\mathrm{obs}} ,z) P(M_{\mathrm{obs}}|M, z)~\mathrm{d}M_{\mathrm{obs}}$ in which the dummy variable is taken to be the derived weak lensing mass. When the cosmology is fixed, $P(\gamma | M_{\mathrm{obs}}, z)$ depends only on the halo profile one assumes to fit the shear profile. Thus it suffices to study the distribution $P(M_{\mathrm{obs}}|M)$, which encapsulates the bias and uncertainty of the derived weak lensing mass. The bias and uncertainties of cluster masses in weak lensing analyses have already been thoroughly studied \citep{Clowe2004, Oguri2005, Corless2007, Meneghetti2010, Becker2011, Oguri2011, Bahe2012}. These results of the mass bias and scatter are often quoted directly in cluster cosmology studies or measurements of cluster scaling relations, albeit without considering the impact of selection effects. Such an omission may be justified as previous studies are mostly based on clusters selected through baryonic properties for which the correlation between scattering in baryonic properties and halo masses is of higher orders. In the context of shear-selected clusters, however, such a correlation is direct and might potentially lead to large bias if we derive $M_{\mathrm{obs}}$ from weak lensing shear profiles, as is commonly done in the literature, because the selection of peaks in mass maps and the calculation of $M_{\mathrm{obs}}$ share the same noise, leading to significant distortions of $P(M_{\mathrm{obs}}|M)$. \replaced{It is precisely the goal of this paper to characterize such a bias.}{\edit1{Such a bias is commonly referred as the Eddington bias \citep{EddingtonBias}, and it is precisely the goal of this paper to characterize this well-known bias in the context of weak lensing surveys.}}

This paper is organized as follows. In Sec.~\ref{sec:method}, we introduce the forward modeling framework and analyses we adopt in order to resemble the mass bias in real observations. Results and discussions are presented in Sec.~\ref{sec:results}. As an immediate application, we discuss in Sec.~\ref{sec:appications} the possibility of resolving the anomalous X-ray property found in shear-selected clusters \citep{Giles2015} using our results. Conclusions are given in Sec.~\ref{sec:conclusions}.

\section{Methodology} 
\label{sec:method}
To investigate bias of weak lensing masses for shear-selected clusters, we create mock halo catalogs and generate observed weak lensing mass maps assuming various survey depths to simulate this issue in existing and future surveys.

In Sec.~\ref{subsec:main_method}, we introduce the basic framework in which we calculate the weak lensing mass maps. Since the final samples are obtained through high $S/N$ cuts which inevitably limit the sample size, in order to achieve high statistical significance, we adopt a hybrid framework, in which the analytical halo model is combined with the uncorrelated large scale structure (LSS) noise from ray-tracing simulations, rather than fully resorting to numerical simulations. This hybrid approach allows us to simulate a large number of halos in a reasonably short time scale. Around each simulated halo, the weak lensing mass map is simulated by adding the uncorrelated LSS noise and the statistical noise to assign a $S/N$ value to each simulated halo. Multiple samples are obtained through adjusting the survey depths to generate the mass map and the $S/N$ cuts.

The treatments presented in Sec.~\ref{subsec:main_method} are mostly \replaced{base}{\edit1{based}} on calculation from first principles. In Sec.~\ref{subsec:triaxial}, we discuss the uncertainties originating from the triaxial nature of halos. Since such an uncertainties are more complex to model from first principles, we rely on statistical relations obtained in previous studies to take them into account. Lastly, in Sec.~\ref{subsec:fit}, we derive masses for clusters in each of these samples through a maximum likelihood estimation. The results are presented in Sec.~\ref{sec:results}.

\subsection{The mock cluster samples}
\label{subsec:main_method}
\begin{figure}[t]
    \centering  
    \epsscale{1.175}
    \plotone{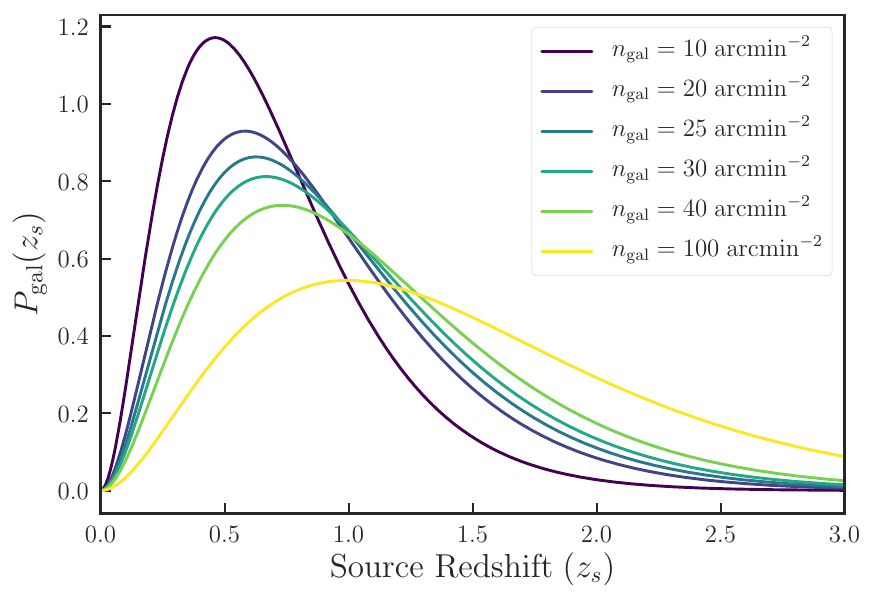}
    \caption{Fiducial models (eq.~\ref{eq:nofz}) of the probability density functions for source galaxies as a function of redshift with source galaxy number densities $n_{\mathrm{gal}} = $ $10,$ $20,$ $25,$ $30,$ $40,$ $100$ $\mathrm{arcmin}^{-2}$ adopted in this paper. \label{fig:nofz}}
\end{figure}

\begin{figure}[t]
    \centering  
    \epsscale{1.175}
    \plotone{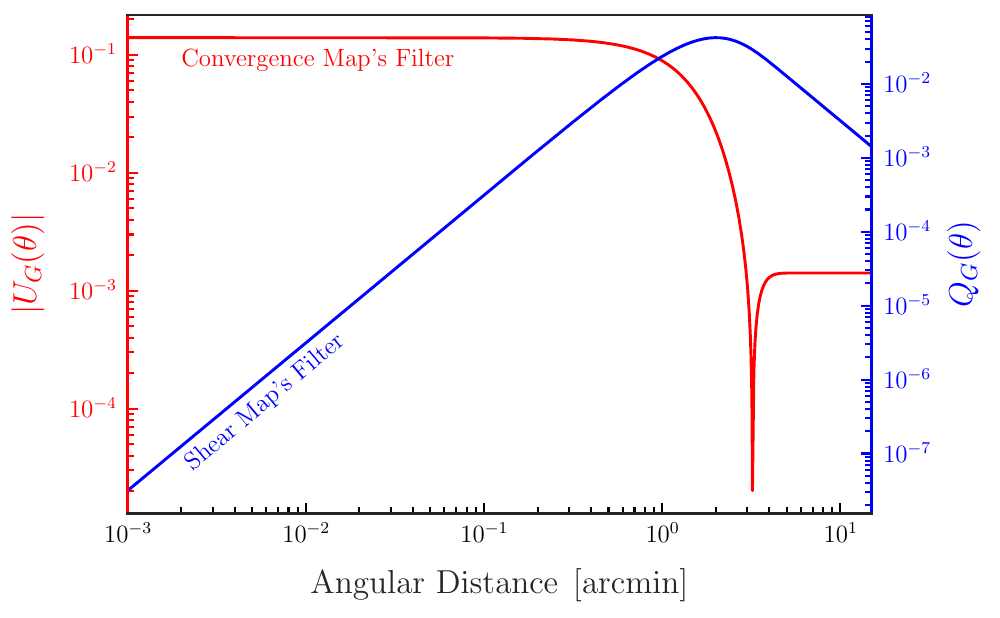}
    \caption{The truncated and compensative Gaussian filter we adopt in this paper. The filter for the convergence (eq.~\ref{eq:Ufilter}) is shown in and the corresponding filter for the tangential shear profile (eq.~\ref{eq:filter}) is shown in blue. \label{fig:filter}}
\end{figure}

The mock samples are set up in the following eight steps, which we describe in turn.
\begin{enumerate}[label=(\roman*), leftmargin=0pt, itemindent=20pt,
labelwidth=15pt, labelsep=5pt, listparindent=0.7cm,
align=left]
    \item We adopt the Nine-Year Wilkinson Microwave Anisotropy Probe (WMAP9) cosmological parameters \citep{WMAP9}\footnote{More specifically, we adopt the parameters given in the column ``+eCMB+BAO+$H_0$'' of Table 4 in \citet{WMAP9}.} as our input cosmology and consider a mock survey with an area of 5000 $\mathrm{deg}^2$. The mock halo sample is then generated as a Poisson realization of the halo mass function \citep{Tinker08}. In our calculations, we consider cluster samples with $M_{200c} \geq 10^{13} h^{-1}M_{\odot}$, where $M_{\Delta c}$ stands for the mass enclosed within the radius $r_{\Delta c}$ where the average density inside is $\Delta$ times the critical density of the Universe. The redshifts of our samples range from $z=0.01$ to $z=1.51$, roughly correspond to the range where optical cluster finding algorithms are able to probe. For our particular experiment, we have created 778651 halos in this step.
    \item For each halo with mass $M=M_{200c}$ and redshift $z=z_{cl}$ in our mock catalog, we assign to it a concentration $c=c_{200c}$ from the distribution 
    \begin{equation}\label{eq:M-c}
        P(c|M, z)= \frac{1}{\sqrt{2\pi}\sigma_{\ln c}}\frac{1}{c}
        \exp\left[-\frac{\big[\ln c-\ln\bar{c}(M, z)\big]^2}{2\sigma_{\ln c}^2}\right],
    \end{equation}
    where the function $\bar{c}(M,z)$ is taken from the mass--concentration relation in \cite{Diemer19} and we adopt $\sigma_{\ln c}=0.25$ that is adequate for the redshift and mass range considered here (B.~Diemer, private communication).
    \item In this step, we compute the tangential shear profile on the sky that is due solely to the halo itself, denoted as $\gamma_{\mathrm{NFW}}(\theta)$, by assuming that the density profile of dark matter halos follows the \citet[][hereafter NFW]{NFW} profile
    \begin{equation}
        \rho_{\mathrm{\mathrm{NFW}}}(r) = \frac{\rho_s}{(r/r_s)(1+r/r_s)^2},
    \end{equation}
    where parameters $\rho_s$ and $r_s$ are computed as
    \begin{equation}\label{eq:mass_def}
        \rho_s = \frac{200\rho_{\mathrm{cri}}(z_{cl})}{3}\frac{c^3}{\ln(1+c)-c/(1+c)};~~
        r_s = \frac{r_{\mathrm{200c}}}{c},
    \end{equation}
    with $\rho_{\mathrm{cri}}(z_{cl})$ being the critical density of the Universe at that redshift. It is then straightforward to compute $\gamma_{\mathrm{NFW}}(\theta)$ as
    \begin{equation}
        \gamma_{\mathrm{NFW}}\left(\theta:=\frac{r_{\mathrm{proj}}}{D_A(z_{cl})}\right)=\frac{\Delta\Sigma(\theta)}{\Sigma_{\mathrm{cri}}} \coloneqq \frac{\bar{\Sigma}(<\theta) - \Sigma(\theta)}{\Sigma_{\mathrm{cri}}},
    \end{equation} 
    where 
    \begin{equation}
        \Sigma(\theta) \coloneqq \int\mathrm{d}x_3\, \rho_{\mathrm{NFW}}\left(\sqrt{x_3^2+r_{\mathrm{proj}}^2}\right).
    \end{equation}
    For these calculations, we use the analytic expressions found in \citet{NFWshear}.
    The reciprocal critical surface density is calculated as
    \begin{equation}
        \Sigma_{\mathrm{cri}}^{-1}(z_{cl}, z_{s}) = 
        \begin{cases}
        0 &\quad\text{if  } z_{s} \leq z_{cl},\\
        \\
        \displaystyle\frac{4\pi G}{c^2}\frac{D(z_{cl})D(z_{cl}, z_s)}{D(z_s)} &\quad\text{otherwise},
        \end{cases}
    \end{equation}
    where $D(z_{cl})$ and $D(z_{s})$ are angular diameter distances between the lens and the observer and the source and the observer, respectively, whereas $D(z_{cl}, z_s)$ is the angular diameter distance between the source and the lens. In reality, since source galaxies are distributed across a wide range of redshifts, one should consider the ensemble average of 
    $\Sigma_{\mathrm{cri}}^{-1}(z_{cl}, z_{s})$. The distribution of source galaxies varies from survey to survey. In order to investigate the impact of the mass bias in different surveys, we set up a one-parameter fiducial model for the distribution, $P_{\mathrm{gal}}(z_s)$, as a function of the number density of source galaxies $n_{\mathrm{gal}}$ following \citet{nofz}
    \begin{equation}
    \label{eq:nofz}
        P_{\mathrm{gal}}(z_s) \propto z_s^2\exp\left(-\frac{z_s}{z_0}\right); ~~z_0 = \frac{1}{3}\left(\frac{n_{\mathrm{gal}}}{30}\right)^{1/3}.
    \end{equation}
    In this paper, we set up experiments for $n_{\mathrm{gal}} = $ $10,$ $20,$ $25,$ $30,$ $40,$ and $100$ $\mathrm{arcmin}^{-2}$ to simulate various past and future surveys. \autoref{fig:nofz} shows $P_{\mathrm{gal}}(z_s)$ for these values of $n_{\mathrm{gal}}$ used in this paper. This profile can approximate the source redshift distribution of real galaxies in e.g., the Subaru HSC survey \citep{HSC} well. The ensemble average of $\Sigma_{\mathrm{cri}}^{-1}(z_{cl}, z_{s})$ is then given by
    \begin{equation}
    \label{eq:surface_cri}
        \langle\Sigma_{\mathrm{cri}}^{-1}\rangle(z_{cl}) = \int_{0}^\infty \Sigma_{\mathrm{cri}}^{-1}(z_{cl}, z_s) P_{\mathrm{gal}}(z_s)~\mathrm{d}z_s.
    \end{equation}
    We caution that the source redshift range of the above integral should be chosen carefully depending on the setup of the analysis. In weak lensing analysis, one often selects source galaxies behind the lens as the foreground galaxies add only noise but not signal. However, as we are using this shear profile to generate a mass map and select clusters from this map, at the point of calculating the mass map, we have no prior knowledge of the redshift of our potential candidates. Thus all the quantities needed to construct the mock mass map have to be integrated over the whole line of sight. This subtlety need to be dealt carefully later when adding the random statistical noise (see step~(vi) below).
\end{enumerate}
\begin{enumerate}[resume*]
    \item The mass map is obtained by convolving the shear map with a filter that maximizes the convergence signal from massive galaxy clusters and suppresses the average LSS contribution. Following \citet{Miyazaki2019}, we choose a truncated and compensative Gaussian filter \citep{Hamana2012}
    \begin{equation}\label{eq:Ufilter}
        U_G(\theta) = 
        \begin{cases}
            \displaystyle\frac{1}{\pi\theta_s^2}\exp\left(-\frac{\theta^2}{\theta_s^2}\right) - U_0 &\quad\text{if }\theta\le\theta_{\mathrm{out}}, \\
            &\\
            0&\quad\text{otherwise,}
        \end{cases}
    \end{equation}
    and the respective filter for the tangential shear profile is 
    \begin{equation}\label{eq:filter}
    \begin{split}
        Q_G(\theta) &= 
        \frac{2}{\theta^2}\left(\int_0^\theta U_G(\tilde{\theta})~\tilde{\theta}~\mathrm{d}\tilde{\theta}\right) - U_G(\theta) \\
        &\hspace*{-5mm}=\begin{cases}
            \displaystyle\frac{1}{\pi\theta^2}\left[1-\left(1+\frac{\theta^2}{\theta_s^2}\right)\exp\left(-\frac{\theta^2}{\theta_s^2}\right)\right] &\;\text{if }\theta\le\theta_{\mathrm{out}}, \\
            &\\
            0&\;\text{otherwise.}
        \end{cases}
    \end{split}
    \end{equation}
    Following \citet{Miyazaki2019}, we choose the smoothing radius $\theta_s = 1\farcm5$ and the truncation radius $\theta_{\mathrm{out}} = 15'$. The parameter $U_0$ in Eq.~\eqref{eq:Ufilter} is introduced to let the filter $U_G$ be of average zero. We plot these filters in \autoref{fig:filter}.
    
    In actual weak lensing observations, the observable for a galaxy cluster is the reduced shear profile $g(\theta)$ derived from the average ellipticity of background source galaxies. Here we adopt the approximation that $g(\theta) \approx \gamma(\theta)$ for all $\theta \leq \theta_{\mathrm{out}}$. This approximation generally does not hold at small radius as the convergence signal is large in the central region of the halo. However, as the reduce shear profile will be convolved with the filter $Q_G(\theta)$ that down-weights the signal from the center as shown in \autoref{fig:filter}, our approximation is justified. Thus the NFW signal $\mathcal{S}_{\mathrm{NFW}}$ assigned to each clusters is
    \begin{equation}
        \mathcal{S}_{\mathrm{NFW}} = 2\pi\displaystyle\int_0^\infty \gamma_{\mathrm{NFW}}(\theta)Q_G(\theta)~\theta\mathrm{d}\theta.
    \end{equation}
 
\end{enumerate}
\begin{enumerate}[resume*]   
    
    \item The contamination from LSS along the line-of-sight is approximated to first order by randomly painting our mock halos onto numerical simulated density field. For this purpose, we utilize the ray-tracing simulations presented by \citet{Sato09}, because these ray-tracing simulations have a relatively high angular resolution with a pixel size of $\sim 0\farcm 15$ and thus are suited for out study focusing on clusters of galaxies. In their work, they performed high-resolution $N$-body simulations and projected particles onto multiple lens planes. 1000 ray-tracing realizations were then constructed through randomly shifting the underlying density field. In each realization, light sources are placed at $z_s \approx$ $0.6$, $0.8$, $1.0$, $1.5$, $2.0$, $3.0$, resulting in an ensemble of six $5\deg \times 5\deg$ weak lensing maps to be weighted according the chosen source redshift distribution. For each halo in our catalog, we randomly choose a realization with a random position on the map. The tangential shear profile contributed from the LSS, denote as $\gamma_{LSS}(\theta)$, is then calculated as the average tangential shear profile around that randomly chosen location and is weighted with the probability distribution $P_{\mathrm{gal}}(z_s)$ defined in Eq.~\eqref{eq:nofz}. In short, 
    \begin{equation}
        \mathcal{S}_{\mathrm{LSS}} = \displaystyle\int_0^\infty \frac{\sum_{z_s\in S}\langle\gamma^{z=z_s}_{\mathrm{LSS}}(\theta)\rangle P_{\mathrm{gal}}(z_s)}{\sum_{z_s\in \mathrm{S}} P_{\mathrm{gal}}(z_s)} Q_G(\theta)~\mathrm{d}\theta
    \end{equation}
    where the angle bracket stands for the circular average in the two-dimensional weak lensing shear map and $\mathrm{S}$ is the set of available source redshift slices.
    
    \item Next we estimate the statistical noise on the weak lensing mass maps. Two major sources of errors in weak lensing observations are the finiteness and the intrinsic ellipticity of source galaxies. We denote the variance of such an observational uncertainty as $\sigma^2_{\mathrm{STAT}}$. \citet{shape_noise} has shown that such a variance can be quantified as
    \begin{equation}
        \sigma^2_{\mathrm{STAT}} = \frac{\sigma^2_{\epsilon}}{2n_{\mathrm{gal}}}\int_{-\infty}^{\infty}|W(\mathbf{k})|^2~\mathrm{d}\mathbf{k},
    \end{equation}
    where $\sigma^2_{\epsilon}$ is the variance characterizing the intrinsic ellipticity of source galaxies and the factor $2$ accounts for the fact that the tangential shear is derived from one out of two components of galaxy ellipticities, $n_{\mathrm{gal}}$ is the surface number density of source galaxies, and $W(\mathbf{k})$ is the Fourier transform of the spatial filter one adopts to smooth the shear signal (eq.~\ref{eq:filter}). In this paper, we adopt $\sigma_{\epsilon}=0.4$. By the Parseval-Plancherel identity and assuming the noise to be Gaussian, the noise due to observational uncertainty is thus:
    \begin{equation}\label{eq:shape_noise}
        \mathcal{S}_{\mathrm{STAT}}\sim\mathcal{N}\left(0, ~\frac{\sigma^2_{\epsilon}}{2n_{\mathrm{gal}}}\int_{0}^{\infty}2\pi|Q(\theta)|^2~\theta\mathrm{d}\theta\right).
    \end{equation}
    In practice, once we select clusters with a sufficiently large $S/N$, we will be able to obtain their redshift \replaced{base}{\edit1{based}} on their optical counterparts. The shear profile will then be measured with background galaxies only, to increase the signal-to-noise ratio. The noise in this shear profile measurement, combined with the noise due to foreground galaxies, is the source of the noise $\mathcal{S}_{\mathrm{STAT}}$ on the mass map. Thus the two random noises cannot be generated independently. Therefore, instead of using Eq.~\eqref{eq:shape_noise} directly, we first generate a random noise profile solely due to the background galaxies, which will be used later in step (viii)
    \begin{align}
        \gamma^{\mathrm{bg}}_{\mathrm{STAT}, i} &\sim\mathcal{N}\left( 0,~\frac{\sigma^2_{\epsilon}}{2N_{\mathrm{bg, i}}}\right); \label{eq:bg_shape_noise}\\
         N_{\mathrm{bg}, i} &= A_i n_{\mathrm{gal}}\int^{\infty}_{z_{cl}}P_{\mathrm{gal}}(z_s)~\mathrm{d}z_s.
    \end{align}
    Here the index $i$ runs over radial bins that are used in creating the tangential shear profile and $A_i$ is the area of the $i$-th bin. In this paper, we bin the shear profile from $0.2$ to $7~h^{-1}\mathrm{Mpc}$ into 20 bins to be consistent with the existing observational work of \citet{Miyazaki2019}. As discussed in the end of step~(iii), $\mathcal{S}_{\mathrm{STAT}}$ should be due to noise across the whole line-of-sight. Thus, in addition to $\gamma^{\mathrm{bg}}_{\mathrm{STAT}}$, we also calculate the noise profile in the foreground
    \begin{align}
        \gamma^{\mathrm{fg}}_{\mathrm{STAT}, i} &\sim\mathcal{N}\left( 0,~\frac{\sigma^2_{\epsilon}}{2N_{\mathrm{fg, i}}}\right); \\
         N_{\mathrm{fg}, i} &= A_i n_{\mathrm{gal}}\int^{z_{cl}}_0 P_{\mathrm{gal}}(z_s)~\mathrm{d}z_s. 
    \end{align}
    Combining the two noise profiles and convolving this with the desired filter, we obtain the final noise in the mass map
    \begin{align}
        \gamma_{\mathrm{STAT}, i} &= \frac{N_{\mathrm{fg}, i}\gamma^{\mathrm{fg}}_{\mathrm{STAT}, i}+N_{\mathrm{bg}, i}\gamma^{\mathrm{bg}}_{\mathrm{STAT}, i}}{N_{\mathrm{fg}, i}+N_{\mathrm{bg}, i}}; \\
        \mathcal{S}_{\mathrm{STAT}} &= \sum_{i}\left(\gamma_{\mathrm{STAT}, i} Q(\theta_i) A_i\right). \label{eq:alternative_shape_noise}
    \end{align}
   One can easily check that $\mathcal{S}_{\mathrm{STAT}}$ generated as in Eq.~\eqref{eq:alternative_shape_noise} follows the distribution of Eq.~\eqref{eq:shape_noise} in the continuous limit.
   \item We are now able to assign to each halo in our catalog an observable $\nu_{\mathrm{obs}}$ as
   \begin{equation}
       \nu_{\mathrm{obs}} = \frac{\mathcal{S}_{\mathrm{NFW}} + \mathcal{S}_{\mathrm{LSS}} + \mathcal{S}_{\mathrm{STAT}}}{\sigma_{\mathrm{STAT}}}, \label{eq:nu_definition}
   \end{equation}
   which represents the signal-to-noise ratio of a peak in the weak lensing mass map for that halo. Eq.~\eqref{eq:nu_definition} is normalized by $\sigma_{\mathrm{STAT}}$ because in observations the noise is estimated from many realizations of weak lensing mass maps with randomly rotated source galaxies, which eliminates the LSS noise but retains only the statistical noise \citep[see e.g.,][]{Miyazaki2019}.
   The shear-selected clusters are identified as those with $\nu_{\mathrm{obs}}\geq\nu_{\mathrm{threshold}}$. We vary the value of $\nu_{\mathrm{threshold}}$ to investigate the mass bias as a function of the selection criterion. Observationally, the redshift is given by their optical counterparts. Most of the state-of-the-art optical cluster finding algorithms are able to produce a pure and complete cluster sample at low redshift and assign to them accurate photometric redshifts. For insatnce, the CAMIRA algorithm \citep{CAMIRA}, adopted by the HSC collaboration for generating the optical cluster catalogs, is able to estimate cluster (photometric) redshifts to high accuracy, $\sigma_{z}/(1+z) \lesssim 0.01 \text{ for } z_{cl} \le 0.8$ \citep{CAMIRA2}. Furthermore, \citet{Miyazaki2019} have shown that peaks with high $S/N$, $\nu_{\mathrm{obs}}\geq 4.7$, found in HSC weak lensing mass maps have optical counterparts almost completely.
   Thus in this paper, we assume that all peaks on the mass map are able to find optical counterparts and can be assigned with a $z_{cl}$ with a negligible error. 
   
   \item For clusters selected in the above step, we also create mock tangential shear profiles that are used to derive weak lensing masses for shear-selected clusters in Sec.~\ref{subsec:fit}. The main signal is the shear profile due to the massive halo itself as described in step~(iii). The only difference is that instead of using all the source galaxies along the line-of-sight to obtain the shear signal, we use only background galaxies as is often done in actual analyses. This setup alters the ensemble average of $\Sigma_{\mathrm{cri}}^{-1}(z_{cl}, z_{s})$ from Eq.~\eqref{eq:surface_cri} to
   \begin{equation}
       \langle\Sigma_{\mathrm{cri}}^{-1}\rangle(z_{cl}) = \frac{ \int_{z_{cl}}^\infty \Sigma_{\mathrm{cri}}^{-1}(z_{cl}, z) P_{\mathrm{gal}}(z_s)~\mathrm{d}z_s }{\int_{z_{cl}}^\infty P_{\mathrm{gal}}(z_s)~\mathrm{d}z_s}.
   \end{equation}
   This continuous signal $\gamma^{\mathrm{bg}}_{\mathrm{NFW}}(\theta)$ is binned into 20 bins in physical scale as discussed in step~(vi). Around the same location chosen in step~(v), a shear profile due to the LSS is calculated using only background sources with $z_s \geq z_{cl}$. The statistical uncertainties only from background galaxies generated in Eq.~\eqref{eq:bg_shape_noise} is added to the profile. The resulting shear profile, $\gamma^{\mathrm{bg}}_{\mathrm{obs}, i} = \gamma^{\mathrm{bg}}_{\mathrm{NFW}, i}+\gamma^{\mathrm{bg}}_{\mathrm{LSS}, i}+\gamma^{\mathrm{bg}}_{\mathrm{STAT}, i}$ $(i = 1, \ldots, 20)$ is the mock observable that we use to derive the weak lensing mass and quantify the mass bias.
\end{enumerate}

\begin{figure*}[t]
    \centering  
    \plottwo{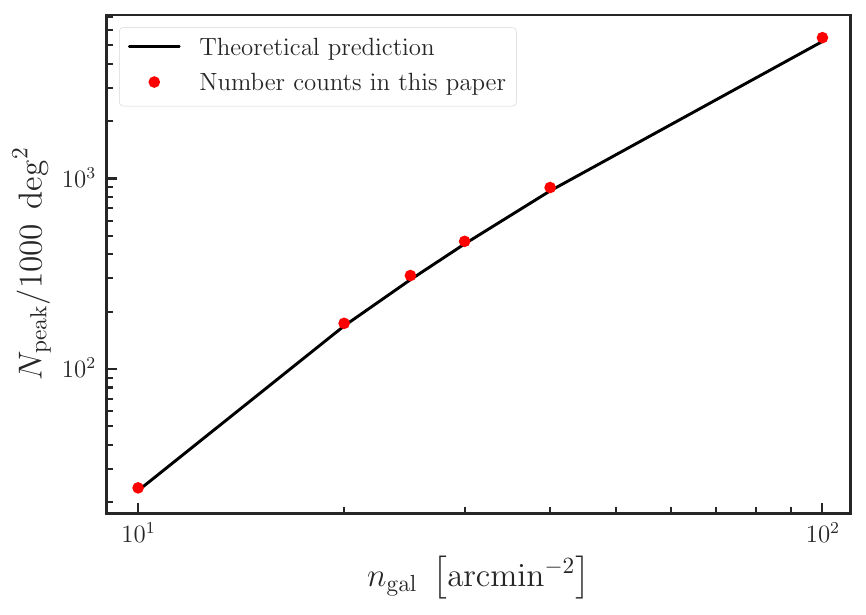}{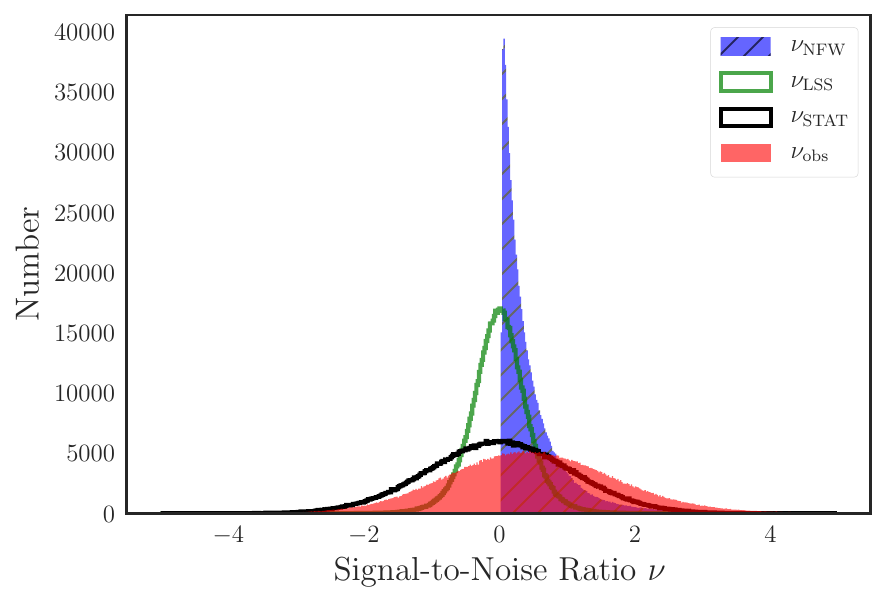}
    \caption{\textit{Left:} Comparison between number counts of shear peaks with $\nu_{\mathrm{threshold}}=4.7$ obtained in this paper and those calculated in Appendix A of \citet{Miyazaki2019}. \textit{Right:} The distributions of $\nu$ for various components that are used to derive the final observed $S/N$ of peaks, $\nu_{\mathrm{obs}}$, for a setup with $n_{\mathrm{gal}}=30~\mathrm{arcmin}^{-2}$. \label{fig:peak_count}}
\end{figure*}

To summarize, we fix a cosmology and change the input parameters $(n_{\mathrm{gal}}, \nu_{\mathrm{threshold}})$ to generate different shear-selected cluster samples with observables $(\nu_{\mathrm{obs}}, z_{cl},$ $\{\gamma^{\mathrm{bg}}_{\mathrm{obs}, i}\}_{i=1,\ldots,20})$ that are to be analyzed in Sec.~\ref{subsec:fit}. \autoref{fig:peak_count} shows some statistics for the catalog we created. The left panel shows the number of clusters that is selected with $\nu_{\mathrm{threshold}}=4.7$ as a function of input $n_{\mathrm{gal}}$ in comparison with the analytic calculation performed in Appendix A of \citet{Miyazaki2019}. The sample sizes increase substantially for deeper surveys, demonstrating the advantages of shear-selected cluster samples in the future. On the right panel, we show the typical level of scattering in the mass map. Since the signals from the halos themselves are buried in the noise except for very massive clusters, a large number of low mass objects can be up scattered into our selections. Thus we can already expect to find a considerable bias in shear-selected samples. 

\added{\edit1{Lastly, we comment on the possible impacts from the specific choice of mass--concentration relation and halo density profile might have on our mock cluster samples. The mass--concentration relation has been heavily studied in the literature, and a number of functional forms, calibrated with cosmological simulations, have been proposed to describe this relation \citep[e.g.,][]{Bullock2001, Maccio2008, Prada2012, Ludlow2016, Diemer19}. Here, we have examined that the distribution of our observable $\nu$ is basically unaffected under different choices of mass--concentration relation, especially for low mass halos. Since we expect the bias to be coming from up-scattered low mass objects, we believe the impact from the choice of mass--concentration relation will be negligible. Meanwhile, in this paper we work with the NFW model to forecast the weak lensing observable due to the simplicity of its lensing properties. However, recent high resolution cosmological simulations have found that the halo density profiles are more accurately described by the functional form proposed by \citet{Einasto1965}, especially at inner radii of halos \citep{Merritt2006, Graham2006, Gao2008, Navarro2010}. We have examined that changing the model will only slightly vary the value of $\nu$ due to the shape of the filter we adopt in this paper. Moreover, the two models nearly coincide with each other at the radius range, $0.3$ to $7~h^{-1}\mathrm{Mpc}$, we considered when generating the shear profile for most of our halos. Thus, we do not expect the choice of halo density profile to affect our derived weak lensing mass either. On the other hand, we note that deviations from the spherical symmetry assumption for both models might induce non-negligible effect on our results. This will be discussed in the following subsection. }}

\subsection{Impact of the halo non-sphericity}
\label{subsec:triaxial}

In reality, dark matter halos are complex and their density profiles deviate from the simple NFW model we assume in \ref{subsec:main_method}. One of the potentially important sources of systematic errors comes from the triaxial shape of real halos \citep{Warren1992}. It has been shown that a spherical symmetric NFW fit to a triaxial halo can lead to an error larger than $30\%$ in mass in some cases \citep{Oguri2005, Corless2007, Meneghetti2010, Becker2011, Oguri2011, Bahe2012}. It may therefore be necessary to take these uncertainties into account.

However, addressing the triaxial nature of the halo density profiles directly is non-trivial. Although analytic triaxial profiles are described for instance in \citet{Jing2002}, statistical properties such as the mass--concentration relation in these frameworks is less well known among the literature. In this paper, we make use of existing studies to characterize these uncertainties in a statistical manner. Given a realistic halo with a mass $M_{\mathrm{True}}$, let $M_{\mathrm{NFW}}$ be the mass in which the corresponding spherical NFW profile best describes the projected surface density of the realistic halo. \citet{Becker2011} and \citet{Bahe2012} have shown that the distribution $P(M_{\mathrm{NFW}}| M_{\mathrm{True}})$ is well characterized by a log-normal distribution. Therefore, we assume
\begin{equation}\label{eq:2Dscatter}
    M_{\mathrm{NFW}} = b_{\mathrm{NFW}}M_{\mathrm{True}}e^{\Delta_{\mathrm{NFW}}},
\end{equation}
where $\Delta_{\mathrm{NFW}}\sim \mathcal{N}(0,\sigma^2_{\mathrm{NFW}})$. Previous studies such as \citet{Becker2011} have also determined these parameters through numerical simulations. In these simulations, both halo triaxiality and the line-of-sight projection contributed to the value of $(b_{\mathrm{NFW}}, \sigma_{\mathrm{NFW}})$. As the line-of-sight contribution has already be taken into account in Sec.~\ref{subsec:main_method}, we need to separate the two when choosing $(b_{\mathrm{NFW}}, \sigma_{\mathrm{NFW}})$. \citet{Becker2011} have shown that the value $b_{\mathrm{NFW}}$ is independent from the line-of-sight integration length and is closed to unity. Meanwhile, regardless of the halo mass or redshift, $ \sigma_{\mathrm{NFW}}$ approaches $0.18$ when the line-of-sight contribution vanishes. This result is roughly consistent with estimates of $\sigma_{\mathrm{NFW}}$ utilizing analytic triaxial profiles presented in \citet{Oguri2005}, \citet{Oguri2009}, and \citet{Hamana2012}. Therefore, in this paper we choose $b_{\mathrm{NFW}}=1$ and $\sigma_{\mathrm{NFW}}=0.18$, which approximately corresponded to a $20\%$ scatter in mass. 

Thus, instead of using the true mass $M_{\mathrm{True}}$ to model the density profile in Eqs.~\eqref{eq:M-c} and \eqref{eq:mass_def}, for each halo, we draw a $M_{\mathrm{NFW}}$ according to Eq.~\eqref{eq:2Dscatter} and repeat all the steps after Step (ii) in Sec.~\ref{subsec:main_method} to forecast the weak lensing signals. In this way, the uncertainties due to the triaxial nature of halos are approximately included. The mass bias from this sample is discussed in Sec.~\ref{subsec:triaxial_result}.

\subsection{Mass estimation for shear-selected clusters}
\label{subsec:fit}
To estimate masses for shear-selected clusters, we perform a maximum likelihood analysis on the cluster shear profile obtained in Sec.~\ref{subsec:main_method} or \ref{subsec:triaxial}. The free parameters in our analysis are the mass $M_{200c}$ and the concentration $c_{200c}$. Here, we adopt a Gaussian likelihood on the observed shear profile $\gamma_{\mathrm{obs}}$
\begin{align}
    &\log \mathcal{L}(\{\gamma^{\mathrm{bg}}_{\mathrm{obs}, i}\}|M_{200c}, c_{200c}) = -\frac{1}{2}\sum_{i,j=1}^{20}\Delta\gamma^{t}_{i}\mathbf{C}^{-1}_{ij}\Delta\gamma_{j};\nonumber\\
    &\Delta\gamma_{i} \coloneqq\gamma^{\mathrm{bg}}_{\mathrm{obs}, i}-\gamma_{\mathrm{NFW}, i}(M_{200c}, c_{200c}),
\end{align}
where $\mathbf{C}$ is the covariance matrix and $\gamma_{\mathrm{NFW}, i}$ are the binned analytic shear profile as described in step~(iii), Sec.~\ref{subsec:main_method}. Here we drop a constant term proportional to $\det \mathbf{C}$ as the covariance matrix is a function of the cluster redshift only, which is fixed in our analysis. The components of the covariance matrix are given by
\begin{equation}
    \mathbf{C}_{ij} = \mathrm{cov}(\gamma_i, \gamma_j) = \frac{\sigma^2_{\epsilon}}{2N_{\mathrm{bg}, i}}\delta_{ij} + \mathbf{C}_{\mathrm{LSS}, ij},
\end{equation}
where $\delta_{ij}$ denotes the Kronecker delta and the term $\mathbf{C}_{\mathrm{LSS}, ij}$ denotes the contribution coming from the uncorrelated LSS \citep{C_LSS}. Here we ignore $\mathbf{C}_{\mathrm{LSS}, ij}$ for computational simplicity as is sometimes adopted in real observations.

\begin{figure}[t]
    \centering  
    \epsscale{1.175}
    \plotone{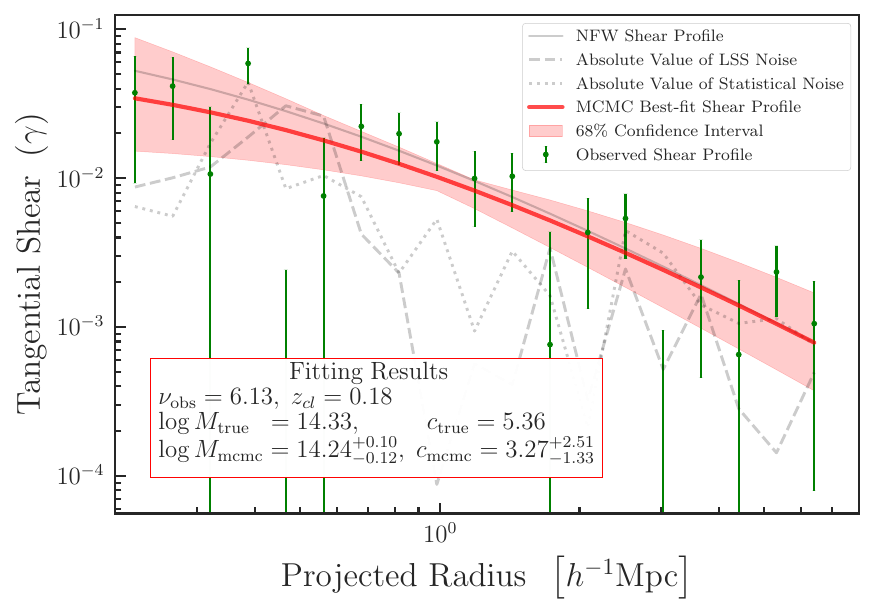}
    \caption{An example of mock tangential shear profile and the resulting fit. Contributions from individual components, $\gamma^{\mathrm{bg}}_{\mathrm{NFW}, i}$ (solid gray), $\gamma^{\mathrm{bg}}_{\mathrm{LSS}, i}$ (dashed), and $\gamma^{\mathrm{bg}}_{\mathrm{STAT}, i}$ (dotted) are shown for a typical cluster in our catalog with $n_{\mathrm{gal}}=30~\mathrm{arcmin}^{-2}$. The solid red line shows the best-fit shear profile with a $68\%$ confidence interval for this particular case. The best-fit parameters are summarized in the inset. \label{fig:fit}} 
\end{figure}

In practice, we sample the physical parameters in logarithmic space using the affine-invariant ensemble sampler \citep{sampler} implemented in the \texttt{emcee} package \citep{emcee}. We assume a flat prior in logarithmic space: $13 \leq \log_{10}\left(M_{200c}~h/M_{\odot}\right) \leq 16$; $0 \leq \log_{10}c_{200c} \leq \log_{10}(20)$. The best-fit parameters are extracted as the 50th percentile of the samples excluding the burn-in period and the uncertainties we quote are the 16th and 84th percentiles. \autoref{fig:fit} shows an example of the mock shear profile $\{\gamma_{\mathrm{obs}, i}\}$ that we assign to a typical cluster and the resulting fit.

\section{Results} 
\label{sec:results}

The important quantity that we study in this paper is the mass bias of weak lensing masses $M_{\mathrm{obs}}$ derived by fitting tangential shear profiles (Sec.~\ref{subsec:fit}) for shear-selected clusters. We define the mass bias $b$ as
\begin{equation}
    b \coloneqq \frac{M_\mathrm{obs}}{M_\mathrm{True}}
    \label{eq:def_massbias}
\end{equation}
In this section, we present the mass bias as a function of different variables by computing the median among the values from clusters within a small bin of the respective variable. The error bars we draw always represent the 16th and 84th percentiles. In order to distinguish the origin of the mass bias, in the following, we first focus on results for the simplest setup in Sec.~\ref{subsec:main_method}, i.e., samples without the effect of non-sphericity discussed in Sec.~\ref{subsec:fit}, to see the bias due solely to the selection together with the statistical and LSS noises. The results with the inclusion of triaxiality will be presented in Sec.~\ref{subsec:triaxial_result}.

\subsection{For spherical halos}
\label{subsec:spherical_result}

\begin{figure*}[t]
    \centering  
    \epsscale{1.175}
    \plotone{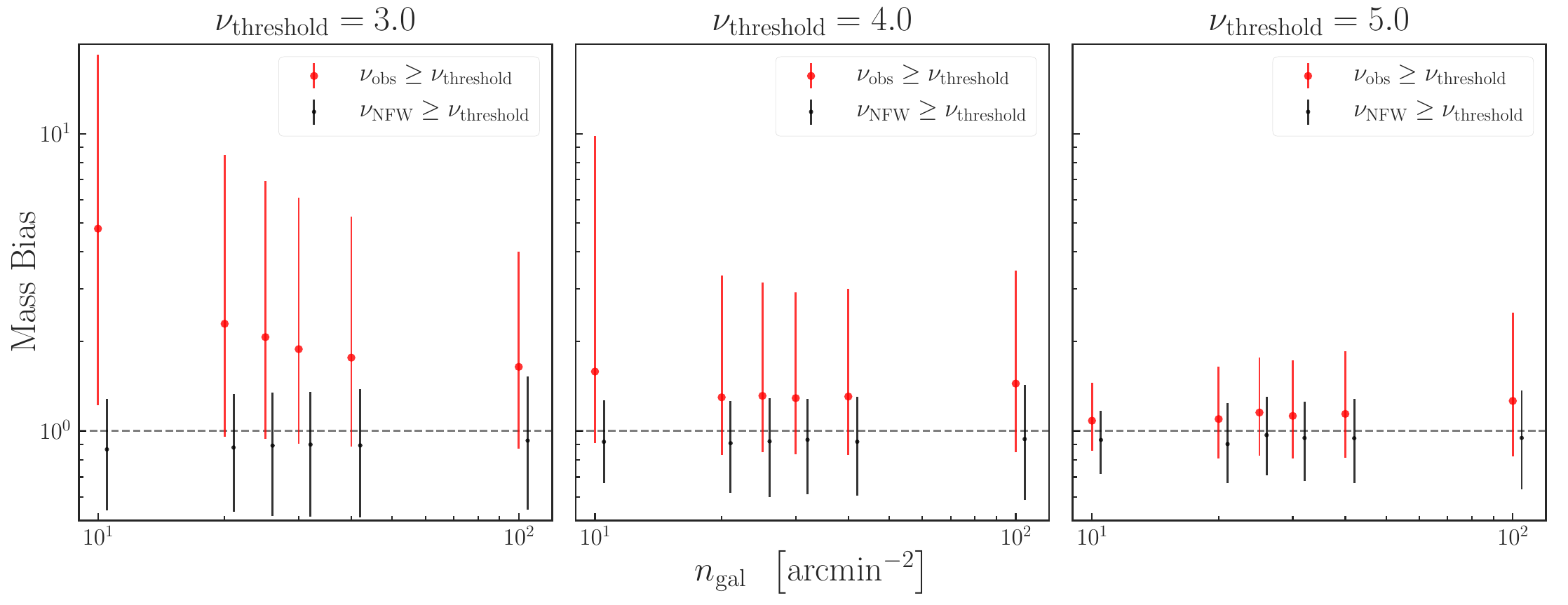}
    \caption{The mass bias (eq.~\ref{eq:def_massbias}) as a function of galaxies number density $n_{\mathrm{gal}}$ under three different selection criterion $\nu_{\mathrm{threshold}}$, $\nu_{\mathrm{threshold}}=3.0$ (left), $\nu_{\mathrm{threshold}}=4.0$ (middle), and $\nu_{\mathrm{threshold}}=5.0$ (right). Large red dots show the results for the shear-selected samples, and small black dots show the results for the corresponding control samples that are selected \replaced{base}{\edit1{based}} on noise-free $\nu_{\mathrm{NFW}}$ instead of $\nu_{\mathrm{obs}}$ to demonstrate that the bias is indeed caused by up-scattering due to the noise. The positions of the control samples are slightly shifted rightwards to avoid overlaps.
    \label{fig:bias_ng}} 
\end{figure*}

To clearly see the effect of up-scatter, for each shear-selected sample, we introduce a corresponding control sample for comparison. Suppose the given shear-selected sample is selected with $\nu_{\mathrm{obs}} \geq \nu_{\mathrm{threshold}}$, the control sample is defined as those clusters with $\nu_{\mathrm{NFW}} \geq \nu_{\mathrm{threshold}}$. The control samples can be viewed as a truly mass-selected sample and hence we expect them to be free from up-scattered contaminants. The mass bias of these two samples will be shown in juxtaposition in the following.

\autoref{fig:bias_ng} shows the bias as a function of the input galaxies number density $n_{\mathrm{gal}}$ under several different selection criteria $\nu_{\mathrm{threshold}}$. \added{\edit1{Here, we clearly see that the bias of each control sample is roughly around one, while the observed samples are all biased high, demonstrating the existence of Eddington bias in shear-selected cluster samples. In \autoref{fig:bias_ng}, the bias of the control sample is slightly below one due to the nature of the noise profile. Although the control sample is selected with a noise-free observable, noise is still included in the shear profile when deriving the weak lensing mass. Therefore, there will be some cases where noise dominants in the observed shear profile. Since the noise profile is flat, only a halo with very low concentration can accommodate such a profile. In these cases, the mass is determined by the average value of the profile, which will usually be lower than the value of the noise-free profile at inner radii due to the steepness of the NFW profile, resulting in a lower best-fit weak lensing mass. Such events will be rarer for larger $\nu_{\mathrm{threshold}}$ or $n_{\mathrm{gal}}$, and indeed, we find the bias to be closer to one in these samples as can be seen in \autoref{fig:bias_ng}.}}

For the actual shear-selected samples, the dependencies of bias on $\nu_{\mathrm{threshold}}$ or $n_{\mathrm{gal}}$ are mainly affected by two factors. Under the same selection threshold, the lower $n_{\mathrm{gal}}$ samples are selecting clusters having larger $M_\mathrm{obs}$ as the noise level is higher in those samples, reducing the value of $\nu_{\mathrm{obs}}$ for a given cluster. Since the slope of the cluster mass function decreases as the mass increases, this leads to a larger fraction of up-scattered objects in the lower $n_{\mathrm{gal}}$ cases. Meanwhile, a higher noise level also allows for larger scattering. At higher $\nu_{\mathrm{threshold}}$, however, these effects become less significant. In reality, the shear-selected samples are usually selected with a high signal-to-noise threshold, e.\,g., $\nu_{\mathrm{threshold}}=4.0$. Thus in the following discussion we will focus on a specific case where $n_{\mathrm{gal}}$ $=30~\mathrm{arcmin}^{-2}$, which roughly corresponds to the depth found in the Subaru HSC survey. 

The relations between the bias and mass or redshift are of the greatest interest in our study, as these are often variables directly used in measuring cluster scaling relations. Understanding the bias as a function of these variables will help us correctly rectify the shape of scaling relations measured through shear-selected cluster samples. 

\begin{figure}[t]
    \centering  
    \epsscale{1.175}
    \plotone{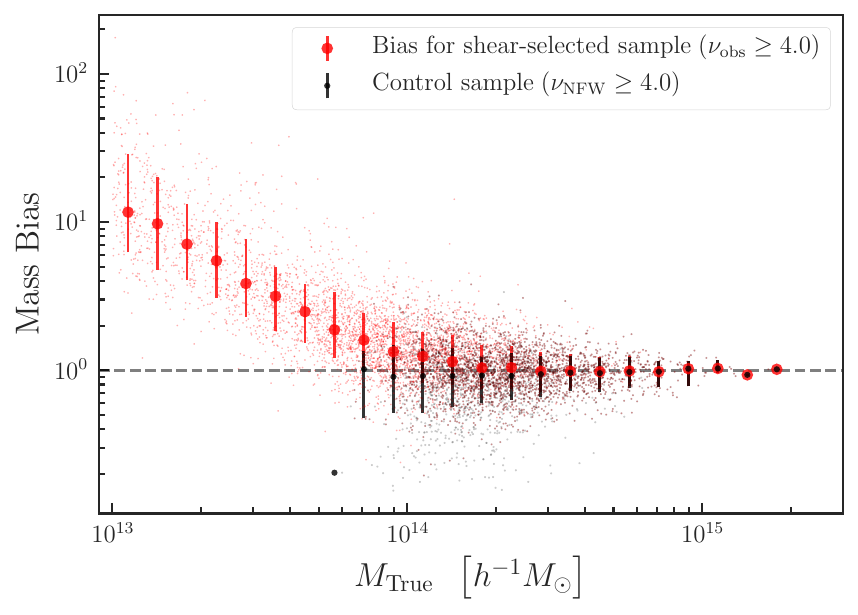}
    \plotone{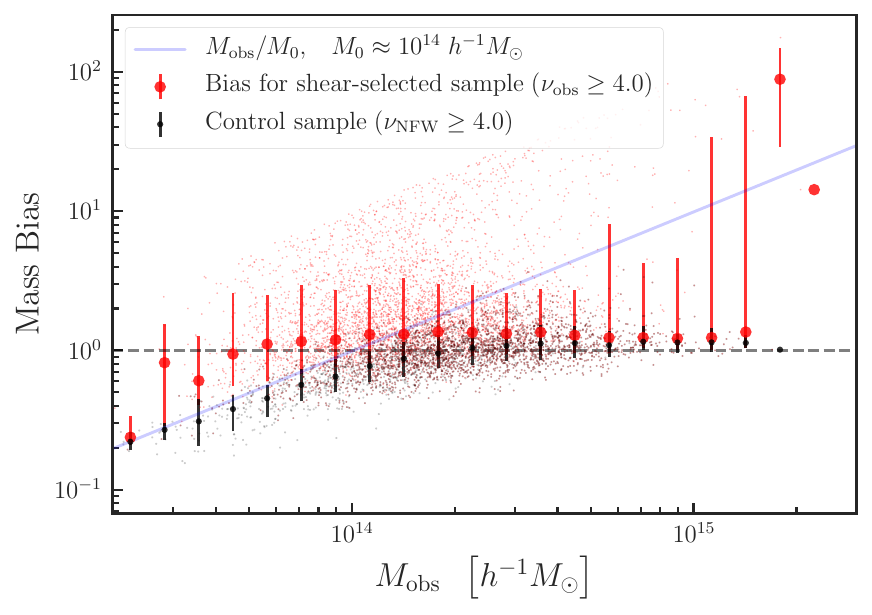}
    \caption{The mass bias (eq.~\ref{eq:def_massbias}) as a function of true mass (upper panel) and the observed weak lensing mass (lower panel). In both panels, large red dots correspond to the bias in the shear-selected cluster sample whereas small black dots denote the bias in the corresponding control sample as in \autoref{fig:bias_ng}. The solid blue line in the lower panel shows the upper bound on the mass bias for the control sample. \label{fig:bias_M}} 
\end{figure}

The cluster mass affects the intrinsic strength of the weak lensing shear signal. Lower mass objects require a larger scatter in order to be selected in the sample, and as a result, the mass bias naturally increases as shown in the upper panel of \autoref{fig:bias_M}. However, as the true mass is not a direct observable, we also show the mass bias as a function of the observed mass $M_{\mathrm{obs}}$. It is clear from the lower panel of \autoref{fig:bias_M} that there are two populations of clusters at a given $M_{\mathrm{obs}}$ range: those that have intrinsic mass in a similar mass range and those that are significantly up-scattered. The situation of up-scattering differs across $M_{\mathrm{obs}}$ but the overall bias seems to be constant within a reasonable range of $M_{\mathrm{obs}}$. We note that in the lower panel of \autoref{fig:bias_M}, the apparent deviation of the mass bias for low $M_{\mathrm{obs}}$ objects in the control sample is artificial. Unlike the shear-selected cluster samples, the selection on $\nu_{\mathrm{NFW}}$ will roughly correspond to a selection on mass, say $M_{\mathrm{True}}\geq M_0$. Therefore, the bias $b = M_{\mathrm{obs}}/M_{\mathrm{True}}$ must be smaller than $M_{\mathrm{obs}}/M_0$. In particular, for $M_{\mathrm{obs}} < M_0$, the bias must be smaller than $1$. For the particular control sample shown in \autoref{fig:bias_M}, $95\%$ of the clusters have mass larger than $10^{14}~h^{-1}M_{\odot}$. Taking this as $M_0$, this upper bound, which is shown in solid blue, clearly explains the trend we see in \autoref{fig:bias_M}. 

\begin{figure}[t]
    \centering  
    \epsscale{1.175}
    \plotone{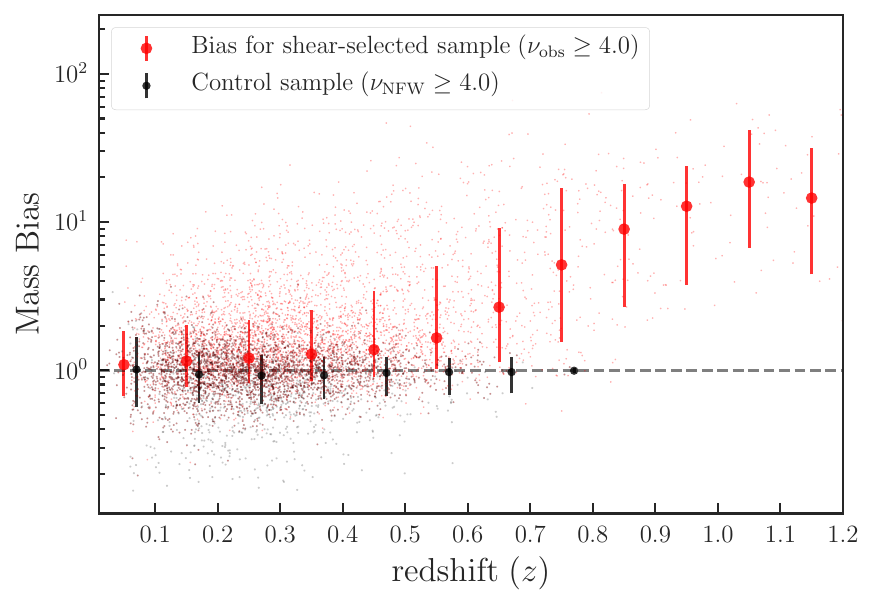}
    \caption{The mass bias (eq.~\ref{eq:def_massbias}) as a function of the cluster redshift. Symbols are same as in \autoref{fig:bias_M}. The redshift positions for the control sample are slightly shifted rightwards to avoid overlaps. \label{fig:bias_z}} 
\end{figure}

The noise in the observed shear profile, on the other hand, is primarily sensitive to the redshift of the lens as the number of source galaxies is more limited at higher redshift. This naturally causes the mass bias to increase as a function of redshift as can be seen in \autoref{fig:bias_z}. \replaced{At high redshifts, we can see that most of the objects will not be selected if not for the large up-scattering.}{\edit1{It is worth noting that in \autoref{fig:bias_z}, the control sample contains no cluster with $z \geq 0.9$, meaning that the cluster candidates observed in the shear-selected sample would not be selected if not for the large weak lensing noise. Therefore, we are actually selecting a sample of purely up-scattered clutsers at higher redshift, and the bias naturally becomes high in these redshift bins.}}

\begin{figure}[t]
    \centering  
    \epsscale{1.175}
    \plotone{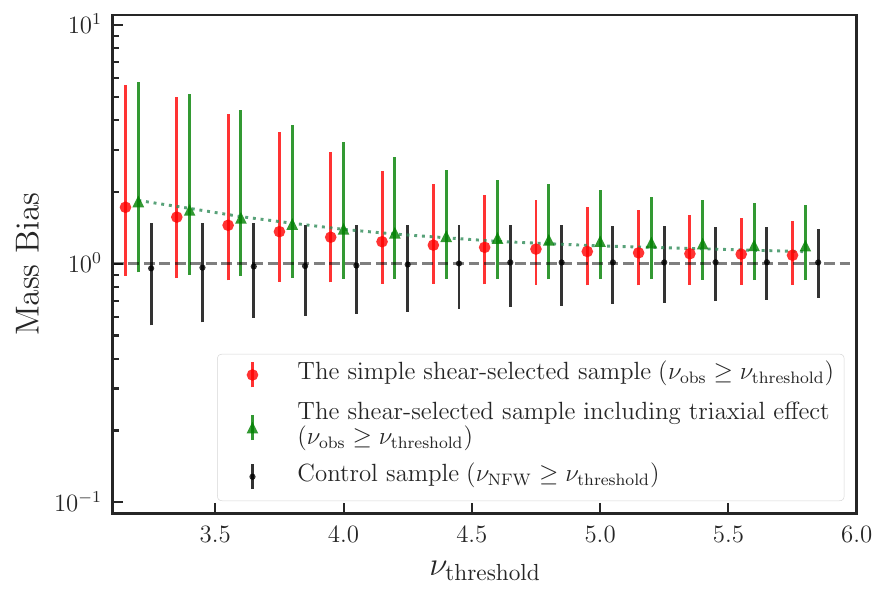}
    \caption{The mass bias (eq.~\ref{eq:def_massbias}) as a function of the selection criterion $\nu_{\mathrm{threshold}}$. The large red dots show the sample constructed in Sec.~\ref{subsec:main_method} whereas the green triangles represent the sample including the modification of halo triaxiality from Sec.~\ref{subsec:triaxial}. The dotted green line shows an quantitative estimate of the results with halo triaxiality (see Sec.~\ref{subsec:triaxial_result} for details). The large red dots and the small black dots are slightly shifted leftwards and rightwards, respectively, from the green tirangles to avoid overlaps. \label{fig:bias_threshold}} 
\end{figure}

For the above results, we show the mass bias in the sample selected with $\nu_{\mathrm{threshold}}\geq 4.0$. The mass bias as a function of this selection criterion is shown in \autoref{fig:bias_threshold}. In real observations, we often modify this lower limit to maximize the number of clusters in the final sample. However, as we lower the $\nu_{\mathrm{threshold}}$, we should also note that we are obtaining a more biased sample.

\subsection{Inclusion of triaxiality}
\label{subsec:triaxial_result}

It is now clear that the selection effect can cause a serious mass bias in a shear-selected cluster sample, even if we assume a spherical halo as an input profile. Here we include additional scatter in mass due to halo triaxiality, which should make our estimate of the mass bias more realistic. We include the effect of halo triaxiality using the methodology  described in Sec.~\ref{subsec:triaxial}. Similarly to the discussions above, we only focus on the fiducial case where $n_{\mathrm{gal}}=30~\mathrm{arcmin}^{-2}$.

An extra scatter in mass from triaxiality can contribute to our results in two ways: it can flatten out the mass function and can induce additional uncertainties in the tangential shear profile. The number density of clusters as a function of scattered mass is equal to the convolution of the original mass function with the distribution function of the induced scattering. Since we are considering a narrow log-normal scatter, the impact on the mass function is minor. For the mock cluster sample considered in this paper and a scatter of $\sigma_{\mathrm{NFW}} = 0.18$, \replaced{the decrease of the number in the low mass end is hardly visible and the number of clusters in the massive end is increased by approximately $5\%$.}{\edit1{the number of clusters in the low mass end ($< 10^{14}~h^{-1}M_{\odot}$) is decreased by roughly $0.1\%$ and the number in the massive end ($> 10^{14}~h^{-1}M_{\odot}$) is increased by approximately $5\%$}} As we are concerned with up-scattering of low mass objects, the portion of clusters that are up-scattered can be treated as the same as before given a same level of scattering. However, the level of scattering can increase due to the extra uncertainty in the shear profile. Within a reasonable range of mass and redshift, a $20\%$ increase in mass will result in a $\sim10\%$ increase in $\nu_{\mathrm{NFW}}$. The effect from such an additional uncertainty, however, is secondary compared to the large statistical noise.

\autoref{fig:bias_threshold} shows the increase in the mass bias after we include the scatter due to triaxiality. Indeed, the increase in the mass bias is more significant at higher $\nu_{\mathrm{threshold}}$ selection as the triaxial scattering starts to become comparable with the statistical noise. Quantitatively, at a given $\nu_{\mathrm{threshold}}=\nu_0$, the level of statistical uncertainty is of order $\sim \nu_0^{-1}$. An additional $10\%$ scatter in $\nu_{\mathrm{obs}}$ will raise the uncertainty to $\sim\sqrt{\nu_0^{-2}+0.1^{2}}$. Equivalently, we can say that the additional scatter shifted the mass bias to the level where $\nu_{\mathrm{threshold}}=\left(\sqrt{\nu_0^{-2}+0.1^{2}}\right)^{-1}$. This estimate is plotted as the dotted green line in \autoref{fig:bias_threshold}. We find that our simple estimate matches \replaced{the mock result}{\edit1{the results on the mock sample}} quite well. For the fiducial case of  $n_{\mathrm{gal}}=30~\mathrm{arcmin}^{-2}$ and $\nu_{\mathrm{threshold}}=4.0$, we find that the mass bias is $\sim55\%$ on average.

\begin{figure*}[t]
    \centering  
    \epsscale{1.175}
    \plotone{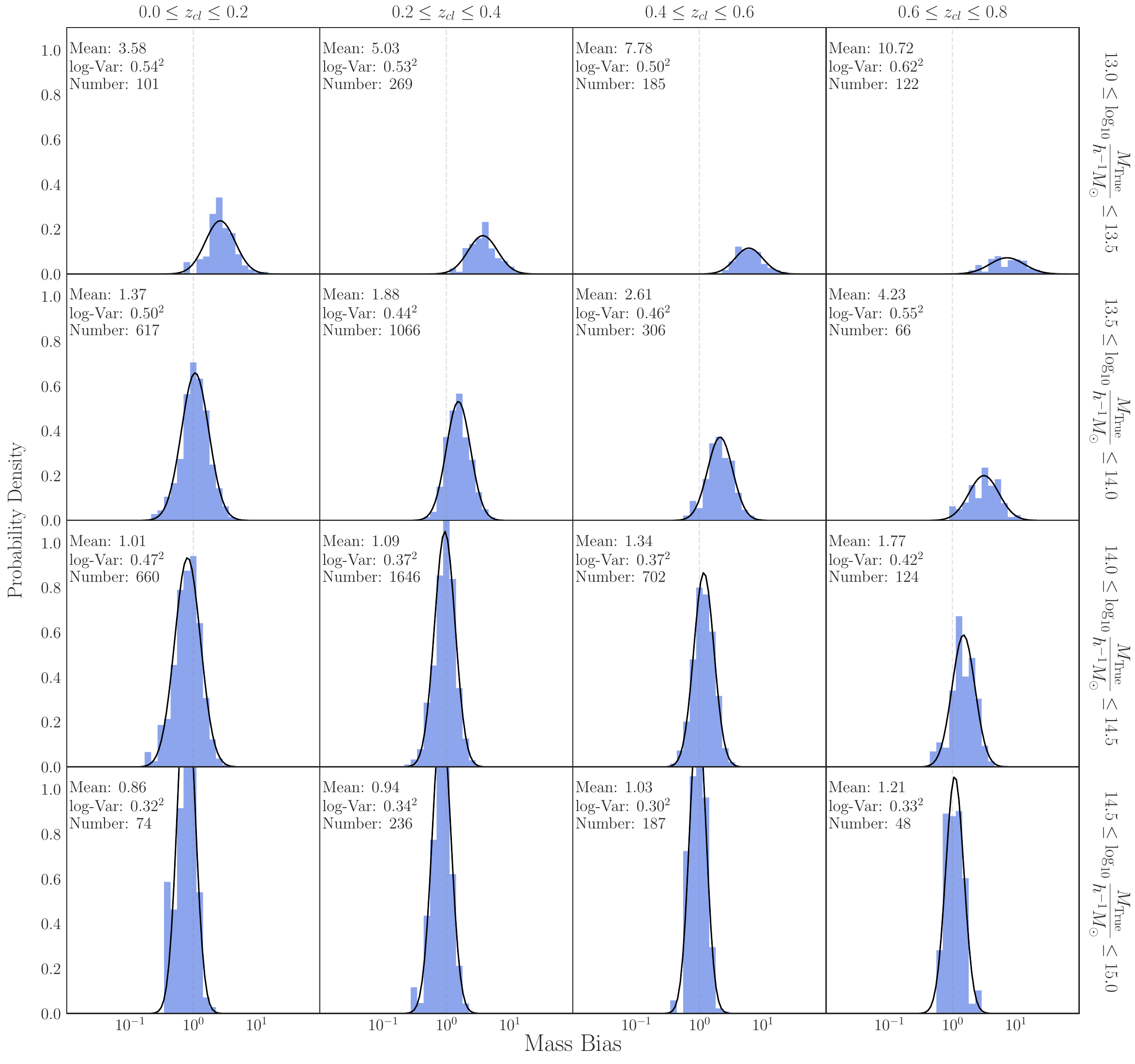}
    \caption{The histograms show the probability density functions (PDF) of the mass bias (eq.~\ref{eq:def_massbias}) in a $4\times4$ grid in redshift and mass. The mean of the distribution and the variance in logarithmic space are quoted on the upper left corner in each panel. In each panel, a log-normal PDF according to these two parameters is over-plotted by a solid line to see how well the distribution is described by the log-normal form. These histograms are based on a $n_{\mathrm{gal}}=30~\mathrm{arcmin}^{-2}$ sample selected with a $\nu_{\mathrm{threshold}} \geq 4.0$ cut. The uncertainty due to halo triaxiality is included. \label{fig:bias_MzScatter}}
\end{figure*}

One of the main purposes of this paper is to serves as a preparatory step for cluster studies utilizing shear-selected cluster samples. So far we are concerned with the average mass bias, but when it comes to constraining scaling relations or cosmological parameters, the knowledge of the scatter in $P(M_{\mathrm{obs}}|M_{\mathrm{True}}, z_{cl})$ is also important. Even if the mass bias is larger in shear-selected cluster samples, tight constraints on cluster scaling relations or cosmological parameters are still possible if the distribution $P(M_{\mathrm{obs}}|M_{\mathrm{True}}, z_{cl})$ is not too dispersed. \autoref{fig:bias_MzScatter} shows the probability density function (PDF) of the mass bias in a grid of $M_{\mathrm{True}}$ and $z_{cl}$ for a sample selected with $\nu_{\mathrm{obs}}\geq 4.0$. We find that these distributions are well-characterized by the log-normal distribution. On the upper left corner of each panel, we show the mean bias and the variance in $\ln{b}$. The corresponding log-normal PDF based on these parameters is over-plotted. Although the mass bias is relatively large, the scatter of the distribution at each redshift and mass is comparable to those without a selection effect \citep{Becker2011, Bahe2012}. Therefore, using shear-selected samples in studies of scaling relations and cosmological parameters might indeed benefit from the absence of uncertainties in semi-empirical mass--observable scaling relations without introducing additional scatter. 

\begin{figure*}[t]
    \centering  
    \epsscale{1.175}
    \plotone{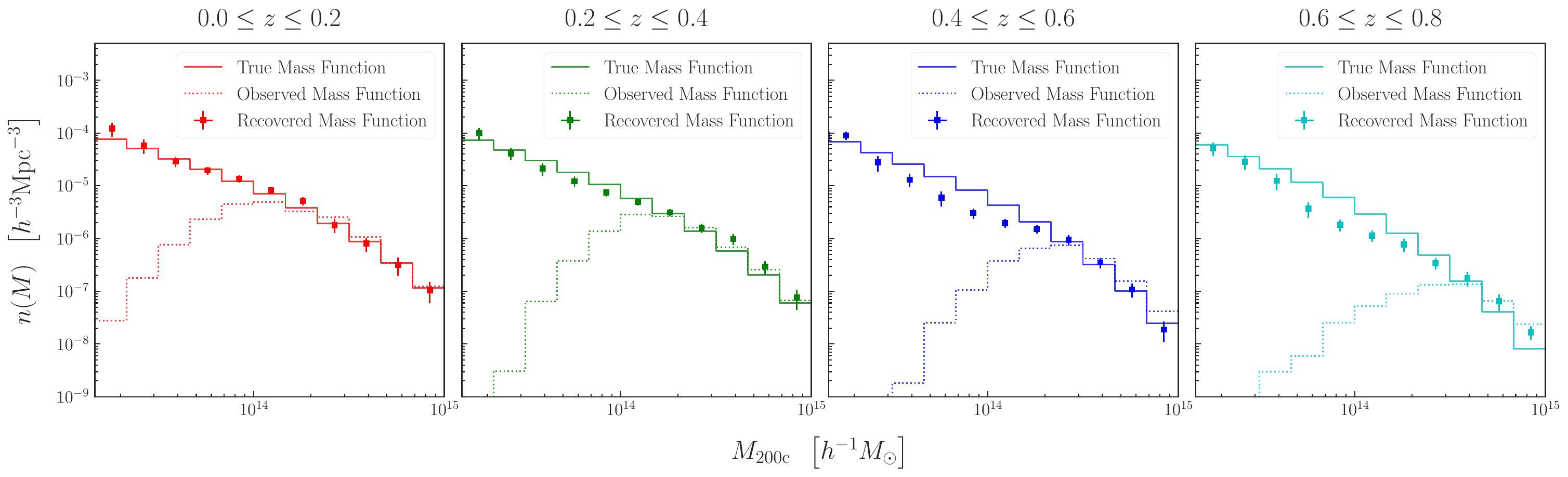}
    \caption{ \edit1{Reconstruction of input mass function from the observed number counts of shear-selected clusters selected in a $n_{\mathrm{gal}}=30~\mathrm{arcmin}^{-2}$ sample with a $\nu_{\mathrm{threshold}} \geq 4.0$ cut. The dotted histograms show the observed number density of clusters in this sample and the reconstructed mass functions are shown in error bars. The true number density in each bin based on the input mass function from \citet{Tinker08} is plotted as the solid histogram.} \label{fig:mass_func}}
\end{figure*}

\added{\edit1{
Finally, as a validation test, we demonstrate that the mass bias and scatter given in \autoref{fig:bias_MzScatter}, together with an analytical description of the selection function, can be used to recover the input halo mass function. Here, we divide the cluster sample selected with $\nu_{\mathrm{threshold}} \geq 4.0$ into four redshift bins as in \autoref{fig:bias_MzScatter}, and count the number of clusters in 12 equally spaced mass bins from $\log_{10}\left(M_{200c}~h/M_{\odot}\right) = 13$ to $15$. The dotted histograms in \autoref{fig:mass_func} show the observed halo mass function in different redshift bins. We can describe this observed mass function with the following integration,}

\edit1{
\begin{align}
    & \displaystyle\frac{\mathrm{d}N}{\mathrm{d}M_{\mathrm{obs}}\mathrm{dz}} (M_{\mathrm{obs}}, z| \nu_{\mathrm{threshold}}) = \int \bigl[P(M_{\mathrm{obs}} | M, z) \nonumber\\
    &\times\frac{\mathrm{d} N}{\mathrm{d} M\mathrm{d} z} (M, z) 
     \times S(M, z | \nu_{\mathrm{threshold}}) \bigr]~\mathrm{d}M, 
     \label{eq:mass_func}
\end{align}
}
\edit1{
where the selection function $S(M, z | \nu_{\mathrm{threshold}})$ is given in Eq. (A9) of \citet{Miyazaki2019}. In practice, we sample the number of clusters in each mass bin $N_{i},~(i = 1, \ldots 12)$ with a flat prior between $0$ to $50000$ while requiring the mass function to be strictly decreasing, we then scatter $N_{i}$ according to the respective log-normal distribution given in \autoref{fig:bias_MzScatter} and multiplied by the average value of the selection function in that bin. We use a Poisson likelihood to compare with the observed number count in each bin to find the most likely set of $N_{i}$ and reconstruct the mass function. The error bars in \autoref{fig:mass_func} show the reconstructed mass function in four different redshift bins while the solid histograms show the average density from the input mass function of \citet{Tinker08}. \autoref{fig:mass_func} shows that we can well reconstruct the input mass function without assuming its functional form at the two lower redshift bins. At higher redshift, the large weak lensing noise makes the number of observed clusters limited even for a $5000~\mathrm{deg}^2$ mock sample, making it more challenging to perfectly recover the input mass function. Meanwhile, deviations from the log-normal distributions assumed for the mass bias at these redshift and mass bins might also contribute to the discrepancy found in \autoref{fig:mass_func}. Overall, however, we think these reconstructions of mass function still give an encouraging evidence that the mass bias derived in this work and the analytic selection function do give a good description of shear-selected cluster samples.}}

\section{Applications}
\label{sec:appications}

The importance of the mass bias reflects on statistical studies carried out with shear-selected cluster samples. \citet{Giles2015} have found that the X-ray luminosity of shear-selected clusters to be lower compare to clusters with similar masses selected in X-ray \citep[e.g.,][]{Mantz2010, Mahdavi2013}. In this Section, we re-analyze their sample with the inclusion of the mass bias found in Sec.~\ref{sec:results} to see whether such a discrepancy can be resolved. We emphasize that our goal here is not to carry out a complete analysis on the scaling relation but just to demonstrate the need for the mass bias in shear-selected cluster samples. 
\subsection{Cluster samples}
First, we briefly summarize the sample presented in \citet{Giles2015}. The cluster candidates were selected as peaks on the $16.72~\mathrm{deg}^2$ mass map in the \textit{Subaru} Weak-Lensing survey \citep{Miyazaki2007} with a $\nu_{\mathrm{threshold}}=3.69$ cut. The survey targeted on $16$ different fields containing X-ray data. Since each field was observed under different seeing condition, the resulting surface number density of galaxies around the cluster candidates was also different. This information is summarized in \autoref{table:G15data}. $28$ of these candidates were spectroscopically confirmed in follow-up observations \citep{Hamana2009}, for which weak lensing masses can therefore be derived. \citet{Giles2015} conducted X-ray studies on $10$ of these clusters, in which eight of them obtained dedicated \textit{Chandra} pointings and the remaining two were observed in the COSMOS field with \textit{Chandra} and in the XMM-LSS field with \textit{XMM}, respectively. The weak lensing and X-ray properties of these 10 clusters are given in \autoref{table:G15data}.

\begin{table}
\centering\scriptsize
\setlength{\tabcolsep}{0.6\tabcolsep}
\begin{tabular}{cccccc}
\hline\hline
Cluster Name & $n_{\mathrm{gal}}$ &$z_{cl}$ & $L_{\rm X}$
& $M_{\rm WL,500}$\\
 & $\left[\mathrm{arcmin}^{-2}\right]$ & & $\left[10^{43}~\mathrm{erg}~\mathrm{s}^{-1}\right]$ & $\left[10^{14}~M_{\odot}\right]$ \\
\hline
SLJ0225.7--0312 & 46.0 & 0.1395 & 7.31$\pm$0.19 & 1.97$^{+0.47}_{-0.47}$ \\
SLJ1000.7+0137  & 37.1 & 0.2166 & 4.04$\pm$0.17 & 2.39$^{+0.46}_{-0.53}$ \\
SLJ1647.7+3455  & 26.4 & 0.2592 & 1.38$\pm$0.12 & 2.00$^{+0.67}_{-0.79}$ \\
SLJ0850.5+4512  & 30.7 & 0.1935 & 0.78$\pm$0.14 & 1.09$^{+0.39}_{-0.43}$ \\
SLJ1135.6+3009  & 29.3 & 0.2078 &      --       & 2.49$^{+0.50}_{-0.56}$ \\
SLJ1204.4--0351 & 23.4 & 0.2609 & 3.43$\pm$0.35 & 1.20$^{+0.50}_{-0.60}$ \\
SLJ1335.7+3731  & 29.6 & 0.4070 & 3.10$\pm$0.71 & 2.79$^{+0.90}_{-1.01}$ \\
SLJ1337.7+3800  & 29.6 & 0.1798 & 0.66$\pm$0.13 & 1.24$^{+0.36}_{-0.39}$ \\   
SLJ1602.8+4335  & 38.0 & 0.4155 & 12.7$\pm$1.14 & 2.66$^{+0.69}_{-0.71}$ \\
SLJ1634.1+5639  & 28.4 & 0.2377 & 0.63$\pm$0.22 & 0.87$^{+0.39}_{-0.49}$ \\
\hline
\end{tabular}
\caption{X-ray and weak lensing properties for the 10 clusters presented in \citet{Giles2015}. The $n_{\mathrm{gal}}$ column represents the average surface number density of source galaxies in the field where the cluster is observed. The $z_{cl}$ column is the cluster spectroscopy redshift obtained in \citet{Hamana2009}. $L_X$ denotes core-excised bolometric X-ray luminosity and $M_{\rm WL,500}$ denotes cluster mass $M_{500c}$ derived with weak lensing. \label{table:G15data}} 

\end{table}

The X-ray luminosity and weak lensing mass scaling relation ($L_X$-$M$ relation) showed that the $10$ shear-selected clusters were X-ray underluminous by a factor of $\sim 2-3$ compared with X-ray selected clusters presented in \citet{Mahdavi2013}. A similar anomaly was also found when comparing with the X-ray samples in \citet{Mantz2010}. \citet{Giles2015} have already presented rough estimates on the mass bias to correct the weak lensing mass. The discrepancies were attributed to miscentering, selection bias, and the triaxial nature of halos. In this paper, we have presented a more rigorous and consistent treatment of these effects. In the following Section, we will adopt our mass bias and examine its impact on the cluster scaling relation.

\subsection{Reanalysis of $L_X$-$M$ scaling relation}
\defcitealias{Giles2015}{G15}
\defcitealias{Mahdavi2013}{M13}

\begin{figure}[t]
    \centering  
    \epsscale{1.175}
    \plotone{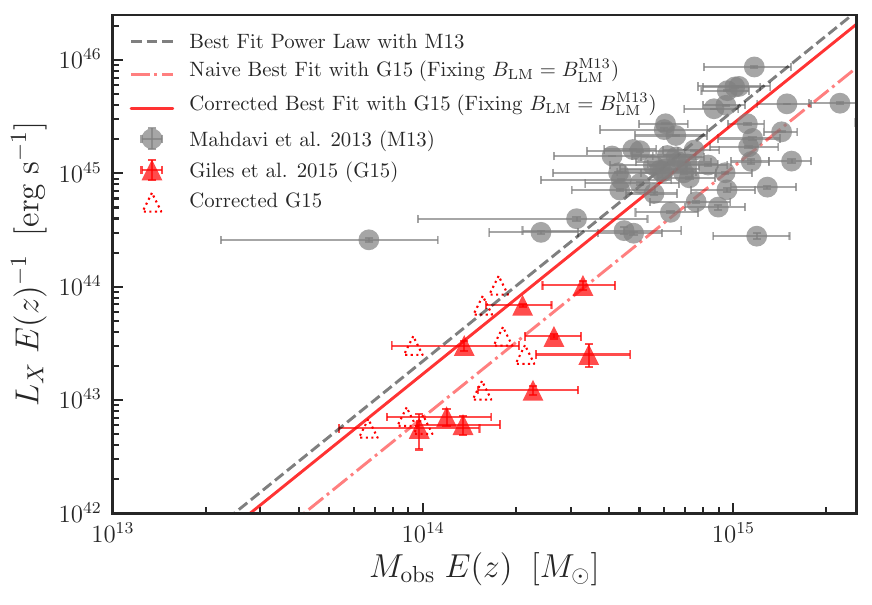}
    \caption{Analyses of the scaling relation between X-ray luminosity $L_X$ and cluster masses $M_{\mathrm{obs}}$ from weak lensing on the dataset from \citet{Mahdavi2013} (grey circles) and \citet{Giles2015} (solid red triangles). A direct power-law-fit to the \citetalias{Mahdavi2013} sample is shown in dashed black line. The best-fit slope $B_{LM}^{\mathrm{M13}}$ is fixed in the analyses on the \citetalias{Giles2015} sample. The dash-dotted red line shows the direct fit to the \citetalias{Giles2015} sample, whereas the solid red line represents the corrected fit including the mass bias. The correction for each individual clusters in \citetalias{Giles2015} is shown in hollow red triangles. \label{fig:Lx-M}}
\end{figure}

Here we follow the procedure in \citet[][hereafter G15]{Giles2015} to analyze the $L_X$-$M$ scaling relation taking account of the mass bias. The scaling relation is described by the following power law
\begin{equation}
    L_X E(z)^{-1} = L_0\left\{\frac{E(z)M_{\mathrm{True}}}{M_0}\right\}^{B_{LM}},
\end{equation}
where $E(z)$ is the dimensionless Hubble parameter and the parameters $(L_0, B_{LM})$ are to be determined from data. The pivot mass $M_0$ is taken to be $2\times10^{14}~\mathrm{M_{\odot}}$. The 10 shear-selected clusters alone cannot provide enough constraining power to fit $(L_0, B_{LM})$ simultaneously. We therefore rely on a larger sample of X-ray-selected clusters presented in \citet[][hereafter M13]{Mahdavi2013} to determine the slope $B_{LM}$ first. The actual data presented here are taken from the erratum \citep{Mahdavi2013Erratum}. The cosmological parameters adopted in the two studies are exactly the same, hence the two samples are compared directly. The best-fit parameters are derived using the orthogonal distance regression \citep{ODR} in order to properly take into account of uncertainties on both variables. The data in \citetalias{Mahdavi2013} gives $B_{LM}^{\mathrm{M13}}=2.21\pm0.37 $ and $L_0^{\mathrm{M13}}= (10.14\pm5.10)\times 10^{43}~\mathrm{erg}~\mathrm{s}^{-1}$. Fixing $B_{LM}=B_{LM}^{\mathrm{M13}}$, the shear-selected sample gives $L_0^{\mathrm{G15}}= (3.23\pm0.77)\times 10^{43}~\mathrm{erg}~\mathrm{s}^{-1}$, corresponding to a $\sim 1.4\sigma$ deviation from the X-ray sample. \autoref{fig:Lx-M} shows the fitting results obtained from the two samples. We find that our fitting result without the correction of the mass bias is consistent with G15.

The discrepancy between G15 and M13 should be to some extent ascribed to the mass bias in shear-selected clusters. As the mass bias depend sensitively on the cluster redshifts but vary little over $M_{\mathrm{obs}}$, we add the following relation when fitting the scaling relation
\begin{equation}
    M_{\mathrm{True}} = b(z_{cl})M_{\mathrm{obs}}.
\end{equation}
The weak lensing mass $M_{\mathrm{obs}}$ for these 10 clusters were derived in \citet{Hamana2009} following a procedure similar to those described in Sec.~\ref{subsec:fit}. Therefore, here we calculate the bias as in Sec.~\ref{sec:method} with a $\nu_{\mathrm{threshold}}=3.69$ selection. The average surface density of source galaxies around the 10 clusters is $\approx 30~\mathrm{arcmin}^{-2}$. Additional scatter due to triaxiality is also included. Since it is customary to use the mass $M_{500c}$ in X-ray studies, we recompute the mass bias changing the mass definition from $M_{200c}$ to $M_{500c}$ in Eq.~\eqref{eq:def_massbias}. We note that there are still two inconsistencies between our analyses in Sec.~\ref{subsec:fit} and those in \citet{Hamana2009}: the probability distribution of source galaxies and the cosmological parameters assumed when deriving the weak lensing mass are different. Nevertheless, if we adopt $n_\mathrm{gal}=30~\mathrm{arcmin}^{-2}$, our numerical calculation shows that Eq.~\eqref{eq:nofz} is almost identical with the functional form assumed in \citet{Hamana2009}. The difference in cosmological parameters induces a deviation in angular diameter distance only by $\lesssim 1\%$ within a reasonable range of redshift. Moreover, the most significant effect from the above mentioned deviations are in the calculation of $\Sigma_{\mathrm{cri}}$, which will be canceled out in the mass bias as we are taking the ratio between masses. Therefore, the usage of our mass bias to this shear-selected cluster sample is justified. 

We then follow the same procedure to re-fit the scaling relation to these 10 clusters. Since this sample was selected with a rather low $S/N$ cut, according to our results, the observed mass for each individual cluster needs to be corrected by a level ranging from $-30\%$ to $-90\%$. The corrected mass for each of these clusters are shown in dotted hollow triangles in \autoref{fig:Lx-M}. With a fixed slope of $B_{LM} = 2.21 \pm 0.37$, the reanalysis on the G15 data gives $L_0^{\mathrm{G15, unbias}} = (7.81\pm1.93) \times 10^{43}~\mathrm{erg}~\mathrm{s}^{-1}$. The tension in the intercept is reduced to $\sim 0.5 \sigma$ after our reanalysis. The result of the new fit is also shown in \autoref{fig:Lx-M}.

We emphasize that our analysis here is far from rigorous. Modern analyses on the scaling relation should include effects such as the Eddington bias, the covariance between variables, and scattering in the mass bias \citep[for a state-of-the-art example, see][]{Dietrich2019}. Although our analysis clearly demonstrated that the mass bias can definitely reduce the tension on the X-ray properties of shear-selected clusters, a more thorough analysis on cleaner and larger shear-selected samples such as those presented in \citet{Miyazaki2019} is needed to draw a definitive conclusion on this issue. 

\section{Conclusions} 
\label{sec:conclusions}

We have studied the bias on masses measured with weak lensing for shear-selected cluster samples, which are constructed from high signal-to-noise ratio peaks in weak lensing mass maps. Since weak lensing mass maps that are used to define shear-selected cluster samples and tangential shear profiles of individual clusters that are used to measure their masses share the same noise, including the statistical noise, the uncorrelated structure along the line-of-sight, and the effect of halo triaxiality, the mass bias on shear-selected cluster samples is expected to be significant. We have quantified this bias accurately using a hybrid approach where the density profile of individual halos is modeled with a simple NFW profile yet high-resolution ray-tracing simulations are used to model the noise originating from the line-of-sight structure. This approach allows us to construct large mock catalogs needed to derive the mass bias accurately.

We have found that the mass bias is indeed significant, particularly for shear-selected clusters constructed from the low number density of source galaxies or from low signal-to-noise threshold of mass map peaks. We have investigated the dependence of the mass bias on different parameters, including true halo masses, cluster redshifts, observed weak lensing masses. We find that the mass bias depends sensitively on cluster redshifts such that clusters at higher redshifts have a larger mass bias. On the other hand, the mass bias depends modestly on observed weak lensing masses. The halo triaxiality has a non-negligible impact on the mass bias, especially when the signal-to-noise threshold or the source galaxy number density is high.

To demonstrate the impact of the mass bias on scaling relation studies, we have applied our result to a shear-selected cluster sample constructed by \citet{Giles2015}. Although it has been claimed, based on analyses of the scaling relation between X-ray luminosity and mass, that shear-selected clusters appear X-ray underluminous by a factor of $\sim 2-3$ compared with X-ray selected clusters, our re-analysis taking proper account of the mass bias for shear-selected clusters has indicated that the discrepancy is significantly reduced. This highlights the importance of the mass bias for shear-selected cluster samples.

Even though the mass bias is large for shear-selected cluster samples, an important advantage of shear-selected cluster samples over other cluster samples is that we can accurately and robustly quantify the mean and scatter of the mass bias. This is because the ingredients needed for calculating the mass bias, including the density profile of clusters, the large-scale structure noise, and halo triaxiality, are fairly well known. In contrast, there are large uncertainties associated with cluster galaxy populations and the physical state of ICM, which implies that it is difficult to robustly model selection functions for cluster samples constructed based on baryonic properties in clusters. One such example is the cool-core bias \citep{Eckert2011}, which modifies posterior distributions of cluster properties such as concentration parameters, asphericity, and the amount of substructures, and hence may induce weak lensing mass bias in a non-trivial way, even though the effect would be smaller than the one studied in this paper. Given the lack of full understanding of baryonic properties in clusters, we face the difficulty in converting these selection functions into the mass bias. Therefore, studies of scaling relations for shear-selected clusters, with the correction of the mass bias as done in this paper, will help identify and quantify known and unknown systematics inherent to scaling relations derived for optical, X-ray, and Sunyaev-Zel'dovich cluster samples. While the current limitation lies in the small number of shear-selected clusters available, ongoing Subaru HSC survey as well as future surveys such as Euclid \citep{Euclid2011} and Large Synoptic Survey Telescope \citep{LSST2009} can construct much larger samples of shear-selected clusters, which will be enormously useful for improving our understanding of cluster scaling relations and hence the reliability of cluster cosmology.

\section*{Acknowledgements}

We thank an anonymous referee for valuable comments and suggestions. KFC and YTL acknowledge support from the Ministry of Science and Technology grants MOST 105-2112-M-001-028-MY3 and MOST 108-2112-M-001-011, and an Academia Sinica Career Development Award (2017-2021).
This work was supported in part by World Premier International Research Center Initiative (WPI Initiative), MEXT, Japan, and JSPS KAKENHI Grant Number JP15H05892 and JP18K03693. Numerical calculations are performed through the \texttt{Python} packages \texttt{Numpy} \citep{numpy} and \texttt{Scipy} \citet{scipy}. Many of the cosmological and astrophysical calculations in this work rely on the routines wrapped up in \texttt{Colossus} \citep{Colossus}. Plots are made available thanks to \texttt{matplotlib} \citep{matplotlib}.

\bibliography{ref}
\listofchanges
\end{document}